\documentclass[prd,preprintnumbers,twocolumn,aps,superscriptaddress]{revtex4-1}
\usepackage{graphicx,hyperref}
\usepackage{dcolumn}
\usepackage{bm}
\usepackage{epsfig,amsmath}
\usepackage{amssymb}
\usepackage[usenames,dvipsnames]{color}

\def \mb{\mathbf}
\def \O{\Omega}

\begin{document}

\title{Intertial Frame Dragging in an Acoustic Analogue spacetime}

\author{Chandrachur Chakraborty}
\email{chandra@pku.edu.cn} 
\affiliation{Tata Institute of Fundamental Research, Mumbai 400005, India}
\affiliation{Kavli Institute for Astronomy and Astrophysics, Peking University, Beijing 100871, China}

\author{Oindrila Ganguly}
\email{lagubhai@gmail.com} 
\affiliation{Institute of Physics, Bhubaneswar 751005, India}

\author{Parthasarathi Majumdar}
\email{parthasarathi.majumdar@rkmvu.ac.in}
\affiliation{Ramakrishna Mission Vivekananda University, Belur Math 711202, India}

\begin{abstract}
We report an incipient exploration 
of the Lense-Thirring precession effect in a rotating 
\textit{acoustic analogue black hole} spacetime. An exact 
formula is deduced for the precession frequency of a
gyroscope due to inertial frame dragging, close to the 
ergosphere of a `Draining Bathtub' acoustic spacetime 
which has been studied extensively for acoustic Hawking 
radiation of phonons and also for `superresonance'. The 
formula is verified by embedding the two dimensional spatial 
(acoustic) geometry into a three dimensional one where the 
similarity with standard Lense-Thirring precession results 
within a strong gravity framework is well known. Prospects
of experimental detection of this new `fixed-metric' effect 
in acoustic geometries, are briefly discussed.
\end{abstract}

\maketitle

\section{Introduction}

A stationary spacetime with angular momentum drags along 
with it local inertial frames, causing a test gyroscope in the spacetime to precess with 
a precise frequency given as a function of the spacetime parameters. 
This is known in the gravity literature as the Lense-Thirring precession 
\cite{lense1918}. Inertial frame dragging 
and the ensuing gyroscopic precession relative to a Copernican frame of
reference are consequences of the `fixed-metric' geometry of a stationary 
spacetime. Until date, most observations and measurements of Lense-Thirring
effect have been carried out only for spacetimes with curvatures 
small enough to justify the weak gravity approximation \cite{kannan2008, iorio2010}. 
It has not been possible so far to directly detect inertial 
frame dragging in regions with very strong gravitational fields as near 
rotating black holes and neutron stars \cite{chandrachur2014a, chandrachur2014b}.

A whole new exciting arena for addressing strong gravity issues has opened up with the discovery by Unruh \cite{unruh1981} 
that acoustic perturbations in the velocity potential of a locally
irrotational, barotropic, inviscid Newtonian fluid behave exactly 
as a minimally coupled massless scalar fields propagating in a \textit{curved}
$(3+1)$ dimensional Lorentzian `acoustic analogue' spacetime. Acoustic analogue 
spacetimes have since been studied extensively to discern laboratory analogues 
of gravitational effects which have so far evaded detection in real spacetime. Unruh's incipient prediction of Hawking radiation of phonons from 
acoustic black hole spacetimes have been the object of study for decades \cite{visser1998,barcelo2005}, culminating in a very recent announcement \cite{steinhauer2016} that correlation between Hawking particles and their partners beyond the acoustic horizon of analogue black holes, in accord with theoretical predictions, have been observed in certain Bose-Einstein condensates. 

Following up on work on Hawking emission of virtual phonons, the acoustic analogue of 
superradiance named as `superresonance' has been quantitatively worked out \cite{basak2003b}. Follow-up work on acoustic superradiance has gone on over the last  decade, with another exciting new announcement made last winter of experimental observation of this effect \cite{torres2016}. One of the key novel predicted features of acoustic superresonance is the possibility that the amplification of the reflected wave may exhibit discreteness characteristic of quantum mechanical vortices. Observation of such discrete jumps in the superradiant scattering coefficient may be a signature of acoustic superradiance. Such theoretical predictions have now made it possible to design in the laboratory simple and robust analogue models of various classes of curved Lorentzian spacetimes in kinematical situations (where the acoustic metric is non-dynamical). Notice that such discrete jumps in the superradiant reflection coefficient is absent for superradiance in physical spacetime.  

In this letter, we explore a new physical effect in 
acoustic geometry - the Lense-Thirring precession of a test gyroscope 
in a rotating acoustic black hole spacetime. Since the phenomenon is 
a familiar one in physical spacetime, and yet has never been observed 
except in its weak field version, acoustic analogue spacetimes afford 
us an arena for a more complete study where strong gravity aspects of 
this precession might be revealed in the laboratory. To this end, we 
shall mostly restrict ourselves to a study of frame dragging close to
the ergosphere of a particular fluid mechanical analogue model of a 
rotating black hole, namely the so-called Draining Bathtub
\cite{visser1998}. In terms of the fluid flow, the ergosphere
characterises a transition surface from a sonic to a supersonic 
flow but sound waves from inside it may still travel outwards, away from the surface.
Such a supersonic region can be 
considered as an acoustic analogue of the ergoregion of a rotating 
black hole, characterised by the reversal in sign of the acoustic
spacetime metric component $g_{00}$. Within this ergoregion, a surface on 
which at every point the normal component of fluid velocity is
inward pointing and greater than the local speed of sound, exists.
This surface then traps all acoustic disturbances within it like an 
outer trapped surface, thus acting as the sonic horizon of an
acoustic black hole. Dragging of local inertial 
frames occurs in the stationary spacetime outside the ergosphere
bounding the ergoregion.

The derivation of the exact Lense-Thirring
precession for four dimensional stationary spacetime has been discussed 
earlier in \cite{chandrachur2014a} (following the textbook \cite{straumannbook}). The derivation of the precession frequency 
to the three dimensional acoustic spacetime situation being discussed here is new, since precession itself is not such an easy idea to visualise in two space dimensions where angular momentum is a scalar quantity. However, we may recall that most elementary treatments of the perihelion precession of Mercury in general relativity is confined to the equatorial plane, and is therefore two dimensional, and yet there has not been any difficulty with its measurement. With this in mind, an exact formula for the precession frequency appropriate to the
Draining Bathtub black hole has been derived here. To ensure that the formula is correct, the three dimensional metric is next embedded into a four dimensional one by simple extension of the geometry along the extra spatial dimension. The known four dimensional formula for the precession frequency is then seen to be completely consistent with what emerges from the extended Draining Bathtub computation. 

The calculated Lense-Thirring precession frequency for the Draining Bathtub acoustic spacetime is seen to enhance rapidly as one approaches the ergosphere from afar. Very close to the ergosphere, it dominates the Kepler frequency (the angular velocity of a test particle/gyroscope which moves along 
a circular geodesic) corresponding to orbiting phonon geodesics, and this provides the clue for its observation. As such, inertial frame dragging in this two space dimensional acoustic spacetime will distort circular geodesics encircling the acoustic black hole, making them encircle with a centre away from the vortex centre. The Lense-Thirring precession would then result in the centre of circular orbits to fluctuate periodically about the centre of the vortex characterising the black hole, with a frequency given by the calculated precession frequency. This frequency should exhibit the same enhancement pattern as one approaches the ergosphere, in accord with our theoretical result mentioned above.

The paper is organised as follows : After a brief review of the Draining Bathtub acoustic spacetime, we present the derivation of our main formula for the Lense-Thirring precession frequency. We then discuss the dependence of this frequency on the radial distance from the acoustic black hole centre, and exhibit its sharp enhancement as one approaches the ergosphere from outside. We then discuss ways that this effect, novel in acoustic gravity, can be observed and the Lense-Thirring precession frequency measured in laboratory experiments. We then conclude with a few remarks on future outlook.

\section{Rotating acoustic black hole}

The acoustic analogue of a rotating black hole spacetime is 
best captured by a planar `Draining Bathtub' flow of an incompressible,
barotropic, inviscid fluid with no global vortex present. The flow is
characterized by the velocity potential 
\begin{align}
\vec v_b = -\frac{A}{r}~\hat{r}+\frac{B}{r}~\hat{\phi}~. \label{dbvel}
\end{align}
Here, $(r, \phi)$ are plane polar coordinates while $A, B$ are constants.
The constraints imposed on the fluid make the background density $\rho_b$ 
position independent which in turn guarantees the constancy of background 
pressure $p_b$ and local speed of sound $c_s$ throughout the fluid. As a 
simplifying measure, we set $c_s = 1$ and ignore an overall constant factor of $\frac{\rho_b}{c_s}$ in the acoustic metric.  The explicit form of the emerging acoustic black hole metric is,
\begin{align}
{ds}^2_{DB}
=
&-
\left(1-\frac{A^2+B^2}{r^2}\right)
{dt}^2
+{dr}^2+r^2~{d\phi}^2 \nonumber \\
&+ 
\frac{2~A}{r}~dr~dt-2B~d\phi~dt ~.
\label{eq:db3metric}
\end{align}
It is clear that this curved analogue spacetime possesses isometries 
that correspond to time translations and rotations on the plane and 
hence, is not only stationary but also axisymmetric. The radius of 
the ergosphere, $r_E$, is determined by the vanishing of 
$g_{00}$: $r_E^2 = A^2 + B^2$. The $2$-surface at $r_H = A$ 
acts as the future event horizon of the sonic black hole because 
just beyond it, the radial component of $\vec v_b$ exceeds the 
local speed of sound. Any linearised fluctuation originating in 
the region bounded by the acoustic horizon is swept inward by the flow.

In the recent literature \cite{chandrachur2014a}, the general 
expression for the angular velocity of precession of a test spin
relative to a Copernican frame of reference has been derived for 
a four dimensional stationary spacetime geometry. 
The Copernican frame, with respect to which the gyroscope precession
rate is given, is not a locally inertial frame. Freely-falling untorqued
gyros cannot precess in such a frame. Rather, a Copernican 
frame \cite{straumannbook} is a local orthonormal tetrad at rest (so moving only in the 
``t" direction determined by the timelike Killing vector of the spacetime) 
and ``locked" to the spatial part of whatever such coordinate system is 
chosen, so that it is ``at rest" with respect to the local inertial frames at infinity.
That is why, a Copernican frame is also called ``axes at rest'' \cite{straumannbook}. As a check on 
our calculations, it makes sense to embed the planar acoustic 
spacetime being considered here, into a three dimensional flow, 
by adding an extra spatial dimension and interpret the result as 
a superposition of an ordinary vortex filament and a line sink:
\begin{align}
{ds}^2_{(3+1)}
=
&-
\left(1-\frac{A^2+B^2}{r^2}\right)
{dt}^2
+{dr}^2+r^2~{d\phi}^2 + dz^2 \nonumber \\
&+ 
\frac{2~A}{r}~dr~dt-2B~d\phi~dt ~.
\label{eq:db4metric}
\end{align}

\section{Frame dragging in acoustic `Draining Bathtub' geometry} \label{sec:framedrag}

Referring to our previous work \cite{chandrachur2014a} for a detailed derivation, the Lense-Thirring precession
($\tilde{\O}$ of Eq.(13) of \cite{chandrachur2014a}) in four dimensional spacetime can be expressed as
\begin{align}
\tilde{\O}
= \frac{1}{2K^2}
*(\tilde K \wedge d\tilde K) \label{ltfreq}
\end{align} 
where,  $\mathbf K$ is the timelike Killing vector field corresponding to the stationarity of the spacetime, and $\tilde{\O}$ is the one-form of Lense-Thirring precession vector $\vec{\O}$. Observe that in two dimensional space appropriate to the acoustic spacetime under consideration, the Lense-Thirring precession frequency ${\O}$, now a spatial scalar, can be expressed by the same formula \autoref{ltfreq}. The above expression is clearly applicable in any $(2+1)$ dimensional \textit{stationary} spacetime.

Specialising to a coordinate basis where $\mathbf K = \mb \partial_0$, 
the corresponding covector $\tilde K$ takes the form:
\begin{align}
\tilde K 
= g_{0 \mu} d \tilde x^\mu 
= g_{00} d \tilde x^0 + g_{0i} d \tilde x^i \label{eq:ktilde}
\end{align}
It follows that,
\begin{align}
d \tilde K &= g_{00,j} d\tilde x^j \wedge d\tilde x^0 + g_{0i,j} d\tilde x^j \wedge d\tilde x^i , \nonumber
\\
(\tilde K \wedge  d \tilde K) &= (g_{00}g_{0i,j} - g_{0i}g_{00,j})  d\tilde x^0 \wedge d\tilde x^j \wedge d\tilde x^i, \nonumber
\\
*(\tilde K \wedge  d \tilde K) &= (g_{00}g_{0i,j}-g_{0i}g_{00,j})*(d\tilde x^0 \wedge d\tilde x^j \wedge d\tilde x^i) \label{eq:kdkcoord}
\end{align}
Using $*(d\tilde x^0 \wedge d\tilde x^j \wedge d\tilde x^i) = \eta^{0ji} = - \frac{1}{\sqrt{-g}} \varepsilon_{ji}$,
we conclude from \autoref{eq:kdkcoord},
\begin{align}
*(\tilde K \wedge  d \tilde K)=\frac{\varepsilon_{ij}}{\sqrt {-g}}
(g_{00}g_{0i,j}-g_{0i}g_{00,j})
\end{align}
Substituting the above in \autoref{ltfreq}, we end up with the following
expression for $\Omega_{(2+1)}$ 
\begin{align}
\Omega_{(2+1)}
= 
\frac{1}{2\sqrt {-g}}\epsilon_{ij} g_{00} \left[\frac{g_{0i}}{g_{00}}\right]_{,j} 
\label{eq:21ltbasis}
\end{align}
To obtain the Lense-Thirring precession rate in Draining Bathtub case, we can easily apply \autoref{eq:21ltbasis} where
$i, j$ take values $1,2$ denoting the two spatial dimensions.
For the line element \autoref{eq:db3metric} of an acoustic black hole,
\begin{align}
\Omega^{DB}_{(2+1)}
&=
- \frac{B~r_E^2}{r^4}
\left[1 - \frac{r_E^2}{r^2}\right]^{-1}
\label{eq:21ltac}
\end{align}

It is clear from \autoref{eq:21ltac} that the the Lense-Thirring precession is more pronounced closer to the ergosphere $r \rightarrow r_E$ than away from it. This is precisely the sort of qualitative behaviour of gyroscopic precession observed in earlier work \cite{chandrachur2014a} on physical stationary spacetimes. 
In the `weak field', i.e., far away $(r >> r_E)$ from the `ergosphere' $|\O^{DB}_{(2+1)}|$ decreases as $1/r^4$:
\begin{align}
|\O^{DB}_{(2+1)}| \simeq \frac{B~r_E^2}{r^4}, \qquad \text{as}\ \ r>>r_E~.
\end{align}
This is also similar to the weak-field approximation that has been used for observation of the gyroscopic precession due to inertial frame dragging in  the Earth's gravitational field arising out of its diurnal rotation \cite{everitt2011}. 

As an extra technical verification that the formula in \autoref{eq:21ltac} correctly gives the Lense-Thirring precession frequency, we use the embedding acoustic 3+1 dimensional geometry given in \autoref{eq:db4metric}, and use our four dimensional formulae \cite{chandrachur2014a}. The Lense-Thirring precession velocity in any $4$ dimensional stationary 
spacetime when expressed in a coordinate basis has the form
\begin{align}
\vec\Omega_{(3+1)}
= \frac{1}{2}\frac{\varepsilon_{ijl}}{\sqrt{-g}} 
\left[ g_{0i,j}
\left(\partial_l - \frac{g_{0l}}{g_{00}} \partial_0 \right)
- \frac{g_{0i}}{g_{00}} g_{00,j} \partial_l
\right]
\label{eq:omega4}
\end{align}
Substituting the metric components from \autoref{eq:db4metric} 
into \autoref{eq:omega4}, we finally obtain the angular velocity 
with which a test gyroscope will precess relative to the frame
$\{\partial_i\}$ in the extended Draining Bathtub spacetime: 
\begin{align}
\vec\Omega_{(3+1)}
= - \frac{B~r_E^2}{r^4}
\left[ 1 - \frac{r_E^2}{r^2} \right]^{-1}
\partial_z 
\label{eq:omegadb}
\end{align}
The formula matches \autoref{eq:21ltac} exactly. With this verification, we can now proceed to consider the physical implications of the gyroscopic precession formula \autoref{eq:21ltac}, vis-a-vis laboratory detection of this effect.

\section{Comparison of Lense-Thirring Precession and background flow rotation}

In this section, we present a comparison of the Lense-Thirring precession angular frequency $\Omega^{DB}_{(2+1)}$ for the Draining Bathtub flow in $(2+1)$ dimensional acoustic spacetime, with the angular velocity of the flow itself, $\Omega$. Observe that the latter is given by,
\begin{align}
\Omega (r) &= \frac{B}{r^2} \label{eq:angvel}
\end{align}
On the other hand, the angular velocity of precession occuring 
due to dragging of local inertial frames has the form,
\begin{align}
|\Omega^{DB}_{(2+1)}(r)|
&=
\left[\frac{B~r_E^2}{r^4}
\left(1 - \frac{r_E^2}{r^2}\right)^{-1}\right]  \nonumber\\
&=\left[\Omega (r) ~ \frac{r_E^2}{r^2}
\left(1 - \frac{r_E^2}{r^2}\right)^{-1}\right] ~, 
 \label{eq:altlt}~
\end{align}
substituting the value of $\Omega (r)=\frac{B}{r^2}$ from \autoref{eq:angvel}. For convenience, we introduce a dimensionless radial coordinate $x \equiv r/r_E$ such that the ergosphere at $r=r_E$ maps to $x=1$ while $r \rightarrow \infty$ corresponds to $x \rightarrow \infty$. Let us also define $\bar \Omega  \equiv |\Omega^{DB}_{(2+1)}|/ \Omega$ which evidently is dimensionless also. \autoref{eq:altlt} then has a simpler appearance:
\begin{align*}
\bar\Omega (x) = \frac{1}{x^2-1}
\end{align*}
It is apparent that when $x=\sqrt 2$, $\bar\Omega = 1$. Thus, at this point, the angular velocity of precession becomes equal in magnitude to the angular velocity of rotation before surpassing it with approach towards the ergosphere ($x \rightarrow 1, ~r\rightarrow r_E$). We denote the corresponding critical radius $r=\sqrt 2~ r_E$ by $r_{c1}$ and $x_{c1} = r_{c1}/r_E = \sqrt 2$. 

\begin{figure}
    \begin{center}
\includegraphics[width=3.0in]{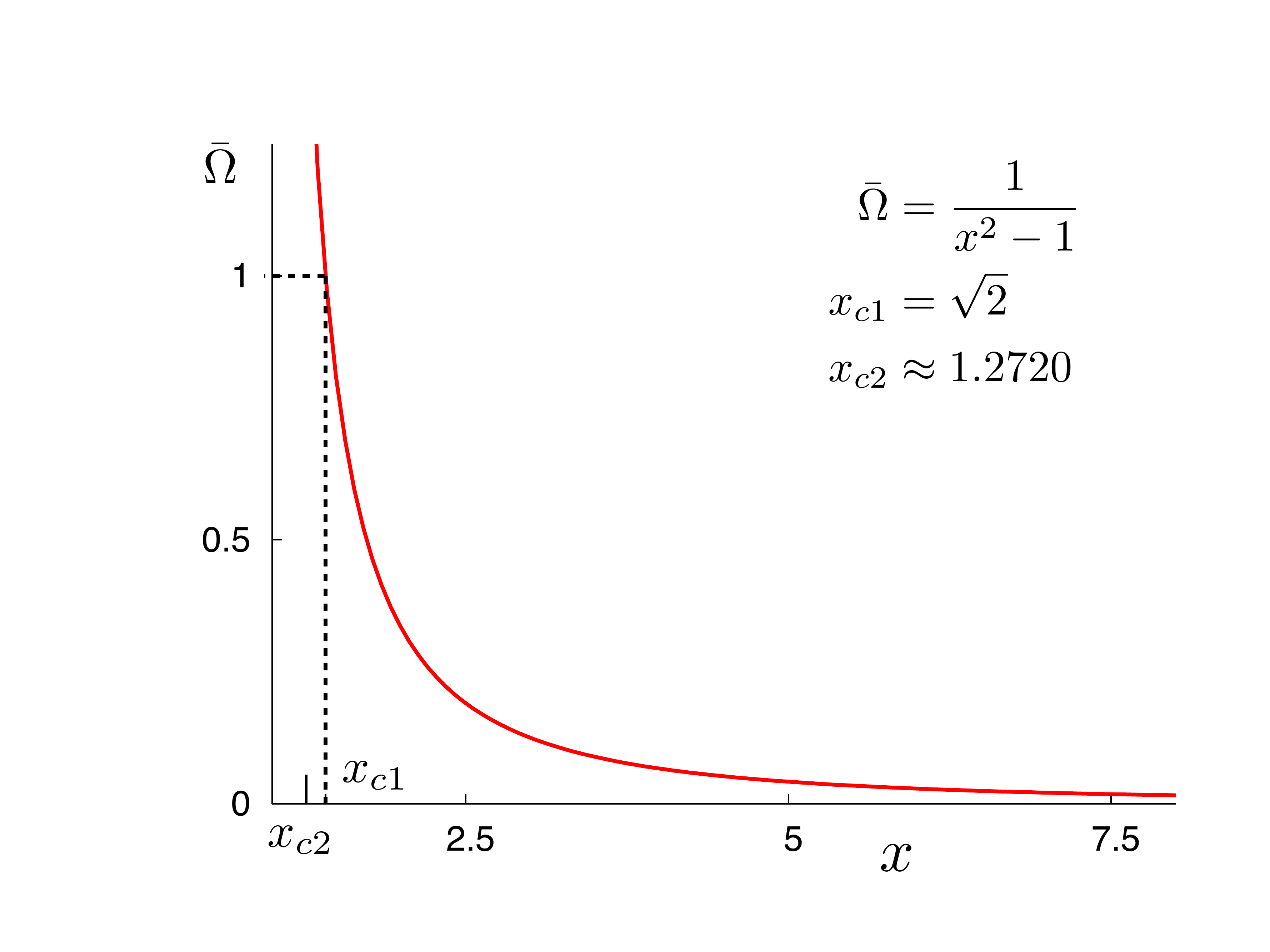}
      \caption{\label{f1}Variation of $\bar\Omega$ with $x$ showing the points $x_{c1}$ and $x_{c2}$}
      \end{center}
\end{figure}

However, it is important to note that the angular velocity of a draining bathtub varies with $r$, unlike in a spinning black hole. Thus, it may also be useful to know at what value of the radial coordinate the magnitude of the precession velocity, $|\Omega^{DB}_{(2+1)}|$, becomes equal  to the angular velocity of rotation of the ergosphere ($\Omega(r_E) = \frac{B}{r_E^2}$). This particular radius denoted by, say, $r_{c2} \approx 1.2720~r_E$ and hence, $x_{c2} = r_{c2}/r_E \approx 1.2720$.  The scenario is graphically represented in Fig. \ref{f1}.


Whether the equality of the Lense-Thirring precession frequency with the angular velocity of the acoustic spacetime at a location outside the acoustic ergosphere signifies a novel \textit{resonance} phenomenon, especially in the context of BEC systems, appears to be an interesting issue worthy of future investigation. Since this equality occurs \textit{outside} the acoustic ergosphere, na\"ively, its connection to `superresonance' is probably peripheral, if at all. However, there may be subtleties which escape us at this point.


\section{Observational Prospects} \label{sec:future}

With Bose-Einstein condensate systems (BEC) providing the most promising experimental situations for more comprehensive studies of superfluidity and other aspects of low temperature physics, it is conceivable that they will also provide the arena for possible observation of inertial frame dragging in an acoustic geometry sensed by an acoustic perturbation on a rotating fluid. Starting with the time-dependent Gross-Pitaevski equation \cite{pethickbook} which follows from the microscopic Bogoliubov equations for the order parameter for a BEC system exhibiting off-diagonal long range order in the mean-field approximation \cite{pitaevskiibook}, the identification of the gradient of the phase of the order parameter with the velocity of the fluid leads to the Euler hydrodynamic equations \cite{garay1999, garay2000, barcelo2005}. 

Assuming that the velocity profile of the Draining Bathtub type is realisable experimentally for a BEC vortex system in the hydrodynamic limit, acoustic perturbations introduced into the system as virtual phonons will be sensitive to such a flow as a rotating analogue black hole, as discussed in the Introduction. Now, rotating Bose-Einstein condensates have been studied extensively \cite{fetter2009}, following earlier work on `BEC gyroscopes' \cite{stringari2001} (and references therein). In these works, the proposed gyroscopes probe the rotating BEC and any precessional motion that the entire BEC system might exhibit. Clearly, this is not adequate for our purpose. To actually observe inertial frame dragging in acoustic geometry, one must consider a gyroscope which rotates freely and probes the behaviour of the virtual phonons.  We assume that it is not difficult to identify observationally part of the BEC vortex for a Draining Bathtub flow as an ergosphere, albeit a small portion of it, as the ergoregion of the acoustic black hole, and an even smaller region as the acoustic horizon. The actual technical difficulties associated with making such identifications are admittedly beyond our range of expertise.

\subsection{Intrinsic Spin of Phonons}

If phonons representing acoustic perturbations inside the fluid have an \textit{instrinsic spin, in addition to their orbital angular momentum}, which is free to precess around the rotation axis of the background flow, one can study such a precession as a \textit{gyroscopic} precession, as a function of its distance from the ergosphere. We have made precise predictions above in \autoref{eq:altlt} for the manner in which the precession frequency must increase as the distance of the phonon from the ergosphere decreases. If the observed behaviour matches with this prediction, it is very likely that one is observing the effect of inertial frame dragging in the acoustic spacetime.  

The notion of an intrinsic phonon spin has been proposed fairly recently by Zhang and Niu \cite{zhang2014} for spin relaxation in ionic crystals exposed to uniform magnetic fields, based on the Raman spin-phonon interaction which is linear in the phonon momentum. Such a momentum dependent interaction was first proposed by Ray and Ray \cite{ray1967} for paramagnetic spin relaxation for Van Vleck two phonon processes, and isotropy (conservation of angular momentum) appears to require existence of this interaction in an essential way. Of course, the importance of such processes depends crucially on the symmetry aspects of the crystal lattice under consideration. It has been argued \cite{zhang2014} that the intrinsic phonon spin is an inherently quantum phenomenon, and it is shown that the spin vanishes in the classical limit. A very clear and pedagogical explanation of this notion of phonon spin has been presented by Garanin and Chudnovsky \cite{garanin2015} where it has been shown that small radius circular shear deformation around the equilibrium points of the elastic medium may produce a non-conserved angular momentum since the stress tensor is clearly non-symmetric in this case. To recover rotational invariance, one may either assume anharmonicity of the elastic medium arising from nonlinearity, or the effect of quantum spin relaxation in the lattice in interaction with phonons. The latter has been argued to induce an intrinsic spin angular momentum for phonons. 

Now, the crucial issue here is: whether BEC systems may model such paramagnetic crystals so that the phenomenon of phonon spin becomes observable in the hydrodynamic approximation in the mean field limit of such condensates. Indeed, the well-known correspondence between Bose-Einstein Condensates and the Bardeen-Cooper-Schrieffer model of superconductivity,  and similar other correspondences  have opened up possibilities of studying phenomena in condensed matter physics far more cleanly within BEC systems confined to certain types of optical lattices \cite{rajibul2017}. We have not been able to discern in the latest literature on BECs as to whether intrinsic phonon spin has indeed been observed in BECs used to model paramagnetic crystals. There is no doubt, however, that within ionic crystals, the phenomenon is now an accepted paradigm, and so it could be a matter of time before it is realised in BEC systems.

There is another possible scenario which can be useful for observation of the phenomenon proposed here, involving \textit{spinor} condensates \cite{pethickbook}, \cite{fetter2009}. Such condensates are likely to have spin wave (magnon) excitations, with possible spin-dependent coupling to phonons. It is not unlikely that because of such interactions, the inertial frame dragging in spinor condensates will result in a flip of the spin wave excitation of the condensate. This spin-flip is a discrete process and ought to be observable in quantum resonance experiments. Once again, there could be a flipping of the magnon spin with a frequency given in terms of the Lense-Thirring frequency. Even though such a scenario is speculative at this stage, since phonon-magnon interactions in spinor condensates are yet to be observed, the one feature that is remarkable is that it may be a discrete process because of quantum dynamics of the spinor BEC. It is similar to the discrete amplification predicted for superresonance in \cite{basak2003b} for BECs or liquid helium. We hasten to add that both these phenomena involving discrete transitions are speculative and have not been observed yet - it is possible that in the linear perturbation theory that we have employed to discern the acoustic spacetime geometry, such discrete transitions cannot possibly be incorporated. 

There is also the issue of using normal fluids like water for observation of the effect of frame dragging in the acoustic geometry. Water has a small viscosity which leads to Lorentz violation \cite{visser1998} in that the acoustic metric may not be extractable in a closed form \cite{ganguly2017}. Nevertheless, it is still possible to demonstrate acoustic superradiance for such fluids, provided the viscosity is within certain limits which also delineates the range of superradiant scattering. The question of how inertial frame dragging may be detected in such fluids is not completely understood yet and will be hopefully addressed in a future publication. 

\section{Conclusions}

Acoustic analogue gravity appears to be richer in terms of observational prospects for non-dynamic spacetimes than actual physical non-dynamic spacetimes. This is so because cold atom condensates present a unique arena for observation of phenomena like Hawking radiation, superradiance and Lense-Thirring precession. These condensate systems involve discreteness because of their inherent quantum mechanical nature; such discreteness has observational prospects for phenomena in acoustic analogue black holes, which one can by no means expect for physical continuum spacetime. It is therefore of utmost importance to investigate thoroughly as to whether the effects discussed in the paper have precise signatures stemming from the unique quantum properties of cold atom condensates.

\section*{Acknowledgements}

We gladly acknowledge S Bhattacharya, K Rajibul Islam and S Sinha for illuminating discussion on Bose Einstein condensates. We also thank an anonymous referee for very incisive comments on the observational aspects presented in an earlier version of the paper.

\bibliographystyle{apsrev4-1}
\bibliography{gravity,an_gr,opub}

\end{document}